\documentstyle[aps, prl]{revtex}
\draft

\begin{document}
\author{Wei-Min Sun$^a$, Xiang-Song Chen$^{a,b}$ and Fan Wang$^a$}
\address{$^a$Department of Physics and Center for Theoretical Physics, Nanjing University, Nanjing 210093, China\\
$^b$Institut f\"ur Theoretische Physik, Universit\"at T\"ubingen, Auf der Morgenstelle 14, D-72076 T\"ubingen, Germany}
\title{Some Problems in Defining Functional Integration over the Gauge Group}
\maketitle 
\begin{abstract}
We find that sometimes the usual definition of functional integration over the gauge group through limiting process may have internal difficulties.
\end{abstract}

Functional integration(or infinite dimensional integration) has become an indispensible tool for the study of modern field theories, especially gauge field theories. There is now no rigorous mathematical definition for this object. Practically one usually defines functional integration as some limit of finite dimensional integrations, as is adopted in most field theory textbooks \cite{Lee}. When the integrand is of quasi-Gaussian type, one can define functional integration rigorously in the framework of perturbation theory, as is advertised in \cite{Faddeev}. These are the definitions for the functional integration encountered in the process of quantising a field theory and they have been studied intensively. In the process of quantising gauge field theory using the Faddeev-Popov technique one also encounters functional integrartion over the gauge group. In \cite{Weinberg} it was utilized to prove the gauge independence of the Green's function of gauge invariant operators. One should define it through limiting processes. In this paper we will show that sometimes this sort of definition may have internal difficulties.

The case we will investigate is $\int_G D\omega \omega^{-1}(x)\partial_{\mu}\omega(x)$, where $G=\prod_xU(1)_x$ is the local gauge group of QED and $\omega(x)=e^{i\theta(x)}$ is the group element of $G$, and we choose the normalisation so that $\int_G D\omega=1$.

Following the usual recipe of defining functional integration we first let the space-time be discretized and replace the derivative $\partial_{\mu}\omega(x)$ by finite differences, after that we will do the finite dimensional integration obtained through the process of discretization and finally we will take the continuum limit.

In the course of replacing $\partial_{\mu}\omega(x)$ by finite differences there are two prescriptions. The first one is to replace $\partial_{\mu}\omega(x)$ by $\frac{\omega(x+\Delta x)-\omega(x)}{\Delta x^{\mu}}$(here we take $\Delta x^{\mu}>0$ for definiteness) while the second one is to replace $\partial_{\mu}\omega(x)$ by $\frac{\omega(x+\Delta x)-\omega(x-\Delta x)}{2\Delta x^{\mu}}$.

In the first prescription 
\begin{eqnarray}
\int_G D\omega \omega^{-1}(x)\partial_{\mu}\omega(x) &=& \lim_{\Delta x\rightarrow 0}\int \prod_{i}d\omega(x_{i})\omega^{-1}(x)\frac{\omega(x+\Delta x)-\omega(x)}{\Delta x^{\mu}} \nonumber \\
&=& \lim_{\Delta x\rightarrow 0}(-\frac{1}{\Delta x^{\mu}}) \nonumber \\
&=& \infty 
\end{eqnarray} 
where we have used the fact that $\int_{U(1)}d\omega \omega=\int_{0}^{2\pi}\frac{d\theta}{2\pi}e^{i\theta}=0$ and $\int_{U(1)}d\omega \omega^{-1}=\int_{0}^{2\pi}\frac{d\theta}{2\pi}e^{-i\theta}=0$. In the second prescription
\begin{eqnarray}
\int_G D\omega \omega^{-1}(x)\partial_{\mu}\omega(x) &=& \lim_{\Delta x\rightarrow 0}\int \prod_{i}d\omega(x_{i})\omega^{-1}(x)\frac{\omega(x+\Delta x)-\omega(x-\Delta x)}{2\Delta x^{\mu}} \nonumber \\
&=& 0 
\end{eqnarray}

One may think that we have found two different ways of defining the functional integration $\int_G D\omega \omega^{-1}(x)\partial_{\mu}\omega(x)$: in the first prescription it is infinity while in the second prescription it is zero. Now we show that both assignments have internal difficulties.

In the first prescription $\int_G D\omega \omega^{-1}(x)\partial_{\mu}\omega(x)=\infty$. We can show that it is in conflict with the property $\int_G D\omega f(\omega)=\int_G D\omega f(\omega^{-1})$ because if this property holds $\int_G D\omega \omega^{-1}(x)\partial_{\mu}\omega(x)$ should be zero:
\begin{eqnarray}
\int_G D\omega \omega^{-1}(x)\partial_{\mu}\omega(x)&=& \int_G D\omega \omega(x)\partial_{\mu}\omega^{-1}(x) \nonumber \\
&=& -\int_G D\omega \omega^{-1}(x)\partial_{\mu}\omega(x)
\end{eqnarray}

In the second prescription $\int_G D\omega \omega^{-1}(x)\partial_{\mu}\omega(x)=0$. We can show that it is in conflict with the property $\int_G D\omega f(\omega_0 \omega)=\int_G D\omega f(\omega)$ because if this property holds $\int_G D\omega \omega^{-1}(x)\partial_{\mu}\omega(x)$ should be infinity: 
\begin{eqnarray}
\int_G D\omega \omega^{-1}(x)\partial_{\mu}\omega(x)&=&\int_G D\omega(\omega_0(x)\omega(x))^{-1}\partial_{\mu}(\omega_0(x)\omega(x)) \nonumber \\
&=& \int_G D\omega (\omega_0^{-1}(x)\partial_{\mu}\omega_0(x)+\omega^{-1}(x)\partial_{\mu}\omega(x)) \nonumber \\
&=& \omega_0^{-1}(x)\partial_{\mu}\omega_0(x)+\int_G D\omega \omega^{-1}(x)\partial_{\mu}\omega(x)
\end{eqnarray}

From the above analysis we see that in this case both assignments are internally inconsistent and the functional integration $\int_G D\omega \omega^{-1}(x)\partial_{\mu}\omega(x)$ cannot be consistently defined. In the following we will construct an example in which this functional integration appears to show that the above discussion is not purely academic.

For an arbitary operator $O(\phi)$, gauge invariant or not, we can always construct an operator $F_O(\phi)=\int_G D\omega O(\phi^{\omega})$, where $\phi$ stands for a gauge or matter field. It can be easily shown that this operator is gauge invariant:
\begin{eqnarray}
F_O(\phi^{\omega_0}) &=& \int_G D\omega O(\phi^{\omega_0\omega}) \nonumber \\
&=& \int_G D\omega O(\phi^{\omega}) \nonumber \\
&=& F_O(\phi)
\end{eqnarray}
Note that in the above proof we have used the property $\int_G D\omega f(\omega_0\omega)=\int_G D\omega f(\omega)$.

Now take $O(\phi)=A_{\mu}(x)$ and see what happens. In this case
\begin{eqnarray}
F_O(\phi)&=& \int_G D\omega A_{\mu}^{\omega}(x) \nonumber \\
&=& \int_G D\omega(A_{\mu}(x)-\frac{i}{e}\omega^{-1}(x)\partial_{\mu}\omega(x)) \nonumber \\
&=& A_{\mu}(x)-\frac{i}{e}\int_G D\omega \omega^{-1}(x)\partial_{\mu}\omega(x)
\end{eqnarray}
If we take $\int_G D\omega \omega^{-1}(x)\partial_{\mu}\omega(x)=0$, as is assigned in the second prescription, we immediately come to an absurd conclusion: the operator $A_{\mu}(x)$ is gauge invariant!

It is not hard to see where we go wrong. The assignment $\int_G D\omega \omega^{-1}(x)\partial_{\mu}\omega(x)=0$ is in conflict with the property $\int_G D\omega f(\omega_0\omega)=\int_G D\omega f(\omega)$, which is essential for the proof of the gauge invariance of the operator $F_O(\phi)$. So it is not strange that we come to a wrong conclusion.

In summary, we find that sometimes the definition of functional integration over the gauge group through limiting process may have internal difficulties.

This work is stimulated by a seminar talk given by Prof. Hung Cheng at Nanjing University in 1999 and his pioneering work in this field \cite{Cheng}. This work is supported in part by the NSF(19675018), SED and SSTC of China, and in part by the DAAD.

\end{document}